\documentclass[journal]{IEEEtran}
\usepackage{times,amsmath}
\usepackage{amssymb}
\usepackage{steinmetz}
\usepackage{pstool}
\usepackage{verbatim}
\usepackage{booktabs}
\usepackage{subfigure}
\usepackage{subcaption}
\usepackage{subfigure}
\usepackage{multirow}
\usepackage{enumerate}
\usepackage{graphicx}
\usepackage{multirow}
\usepackage{subfig}
\usepackage{adjustbox} 
\usepackage[utf8]{inputenc}
\usepackage{tikz}
\usetikzlibrary{arrows, positioning, shapes, decorations.pathreplacing}

\usepackage{float}

\usepackage{mwe}
\usepackage{bigints}
\usepackage{adjustbox}
\usepackage{booktabs}
\usepackage{pdflscape}

\usepackage{MnSymbol}
\usepackage{stfloats} 
\usepackage[table]{xcolor}
\usepackage[square, comma, sort&compress, numbers]{natbib}
\usepackage{nohyperref}
\usepackage{algorithm,algorithmic}
\usepackage{epstopdf}
\usepackage{mathtools, cuted}
\graphicspath{{Figures/3xmulti/}{Figures/3xbinary/}}
\usepackage{algorithm}
\usepackage{algorithmic}
\usepackage{xcolor} % Include xcolor package
\usepackage[table]{xcolor}
\definecolor{lightgray}{rgb}{0.9, 0.9, 0.9} % Define
\definecolor{lightgreen}{RGB}{180, 238, 180} 
\definecolor{green2}{RGB}{100, 238, 100} 
\definecolor{lightred}{RGB}{255, 182, 193}
\definecolor{lightyellow}{RGB}{255, 255, 100} 
\definecolor{lightlightyellow}{RGB}{255, 255, 220} 

\usepackage{rotating}
\usepackage{placeins}

\newcommand{\secref}[1]{Section~\ref{#1}}

\begin{document}

\title{{Neuroscience of Improvisation Across Music and Other Domains: A Survey and Outlook}}
%\title{Improvisation Neuroscience: state of the art, gaps and outlook}
%\title{Neural Dynamics of Improvisatory Brain: A Systems Neuroscience Survey}
%\title{BCI for Neural Dynamics of Improvisatory Brain: A Systems Neuroscience Survey}
%\title{The Improvising Brain: A Systems Neuroscience Framework for Spontaneous, Adaptive Performance}

\author{
\IEEEauthorblockN{
Levina Perzhilla\IEEEauthorrefmark{1},
Muhammad Mahboob Ur Rahman\IEEEauthorrefmark{1}, {\it Senior member, IEEE}, Tareq Y. Al-Naffouri\IEEEauthorrefmark{1}, {\it Fellow, IEEE} }

%\IEEEauthorblockA{\IEEEauthorrefmark{1} Computer, Electrical and Mathematical Sciences and Engineering Division (CEMSE), King Abdullah University of Science and Technology, Thuwal 23955, Saudi Arabia\\ 
%\IEEEauthorrefmark{1}\{muhammad.rahman,tareq.alnaffouri\}@kaust.edu.sa }
\thanks{Authors are affiliated with Computer, Electrical and Mathematical Sciences and Engineering Division (CEMSE), King Abdullah University of Science and Technology, Thuwal 23955, Saudi Arabia. Email: \{muhammad.rahman,tareq.alnaffouri\}@kaust.edu.sa. \\
The research reported in this publication was supported by funding from King Abdullah University of Science and Technology (KAUST) - KAUST Center of Excellence for Smart Health (KCSH), under award number 5932. \\
}
}

\maketitle
\begin{abstract}
Improvisation (or Improv), the real-time generation of novel, context-appropriate action, occurs in music, dance, theater, sport, education, therapy, and professional practice. Its neuroscience, however, has been studied almost exclusively for music, where most quantitative neuroimaging evidence lies. A PRISMA-style search done by authors quantifies this imbalance: of 111 identified studies, 70.3\% concern musical improvisation or general psychometric creativity. This explains why earlier surveys are either music-centered or confined to a single construct. This survey addresses that gap in two ways. First, for music, it organizes findings from functional magnetic resonance imaging (fMRI), electroencephalography (EEG), functional near-infrared spectroscopy (fNIRS), hyperscanning, neurofeedback, and brain stimulation, and reports transparently where the evidence agrees (e.g., experience-dependent changes in frontal engagement, increased frontal alpha and theta power) and where it is heterogeneous (e.g., the direction of lateral prefrontal activity, hemispheric lateralization, the interpretation of alpha power). Second, it extends the review to dance, theater, sport, education, therapy, and professional practice, where behavioral, qualitative, and simulation-based studies predominate. Overall, the current evidence does not point to a universal neural signature of improvisation within music domain or other domains. We conclude with an outlook on wearable EEG--fNIRS sensing and closed-loop brain--computer interfaces for musical improvisation, the domain with sufficient evidence, formulated as an expertise-gated control problem and discussed as a possible complement to deliberate practice, and we list the evidence needed for other domains.
\end{abstract}


\section{Introduction}
\label{sec:intro}

Improvisation is a core feature of human performance, enabling individuals to generate novel and contextually appropriate responses in real time. Unlike rehearsed or pre-planned actions, improvisation emerges in dynamic and often unpredictable environments where flexibility, creativity, and rapid adaptation are required. Across domains---whether a jazz musician weaving an unplanned melodic line, an actor responding spontaneously to a co-performer, a dancer creating movement without choreography, or an athlete adjusting tactics during competition---improvisation functions as a powerful mechanism for effective, resilient, and often highly creative performance \cite{berkowitz2008generation, berkowitz2010cognitive}. Beyond the arts and sports, improvisation plays an equally important role in everyday life, shaping how individuals respond to unexpected problems, navigate social interactions, and make decisions under incomplete information \cite{sawyer2001creating}.

Although improvisation is often discussed alongside related constructs such as divergent thinking, convergent thinking, and creativity, these concepts are hierarchically distinct and reflect different modes of cognition and action. Creativity is a broader construct that encompasses the full process by which ideas or actions become both novel and appropriate, typically integrating divergent generation with convergent evaluation and refinement. Divergent thinking refers to a generative cognitive process that emphasizes the production of multiple, varied ideas from a given starting point and is commonly used as an experimental proxy for creative potential. Convergent thinking is the reduction of possibilities toward a single correct or optimal solution, relying heavily on goal-directed control, rule adherence, and evaluative mechanisms. Improvisation occupies a unique position within this hierarchy, i.e., rather than separating generation and evaluation into sequential stages, it represents an embodied, time-critical instantiation of creativity in which divergent and convergent processes are tightly interwoven and enacted simultaneously through real-time perception–action coupling.

This distinction is particularly evident in professional and high-stakes contexts where improvisational capacity is essential.
%Professional contexts also rely heavily on improvisational capacity. 
Teachers adjust lessons in real time to meet classroom needs \cite{sawyer2004teaching}, surgeons must respond to intraoperative irregularities \cite{crossan1996organizational}, and emergency responders continually generate strategies under high pressure and uncertainty \cite{miner2001organizational}. In all of these settings, improvisation enables adaptive performance where rigid planning would be ineffective or impossible. The triggers for improvisation are multifaceted: some are situational, such as environmental unpredictability or interactive group behavior \cite{nagengast2011uncertainty}; others are dispositional or cognitive, including creativity, openness, and tolerance for ambiguity \cite{beaty2015creativity, sawyer2011explaining}. Critically, the fluency and stability of improvisation are profoundly shaped by expertise, as highly trained practitioners rely on deep internalized knowledge and motor repertoires that allow spontaneous action without the burdens of effortful step-by-step processing \cite{pressing1988improvisation, biasutti2015improvisation}.

The consequences of improvisation are substantial. Improvisational ability supports creative idea generation, facilitates entry into flow states \cite{csikszentmihalyi1990flow}, enhances resilience under pressure, and drives innovation across artistic, athletic, and professional domains \cite{montuori2003organisational, kauhanen2020creativity}. Yet improvisation also carries inherent risks: spontaneity introduces uncertainty, which can produce errors, breakdowns, or coordination failures \cite{barrett1998creativity}. Effective improvisation lies in between generative freedom and constraint, spontaneity and control.

Despite its prevalence, the scientific study of improvisation remains deeply fragmented. Existing research typically isolates one component of the phenomenon---creativity, spontaneity, flow, decision-making under uncertainty, or motor adaptability---without treating improvisation as a unified construct \cite{beaty2015creativity, biasutti2017teaching}. Domain silos further limit integrative understanding: musical improvisation has been disproportionately studied, while sports, acting, and dance have received comparatively limited empirical attention \cite{berkowitz2010cognitive, Thompson2020}. Differences in methodology exacerbate this fragmentation. Neurophysiological findings are difficult to compare across studies employing functional magnetic resonance imaging (fMRI) that measures blood-Oxygen-level-dependent (BOLD) dynamics, electroencephalogram (EEG) that measures oscillatory activity, or hyperscanning that measures inter-brain coupling. These disciplinary and methodological boundaries obscure the possibility that improvisation may reflect a coherent, cross-domain neurocognitive system.

{Within music, studies of expert performers report diverse patterns. Some fMRI studies describe reduced activity in the dorsolateral prefrontal cortex (DLPFC, a region associated with deliberate control and self-monitoring) together with increased activity in the medial prefrontal cortex (mPFC, associated with self-generated thought), an account often labeled \textit{transient hypofrontality} \cite{Limb2008, liu2012neural, dietrich2004flow}; others report engagement of premotor and lateral prefrontal regions \cite{Beaty2015rev, LuYao2017}. EEG studies report increased frontal alpha and theta power \cite{Lopata2017, Sasaki2019}, and experience appears to reduce frontoparietal executive activity \cite{pinho2014connecting}. Outside music, neural recordings are rare. We therefore report the evidence domain by domain and are transparent about where it is heterogeneous, rather than imposing a single unifying account.}

\subsection{Contributions}
\label{sec:contrib}

{This article surveys the neuroscience of improvisation across music, dance, theater, sport, education, therapy, and professional practice. Because the bulk of the quantitative neuroimaging literature lies in music, musical improvisation receives the major focus, followed by an engineering outlook for music. The other domains, in which behavioral and qualitative studies dominate, are covered subsequently. Throughout, we report the evidence transparently, including where findings are heterogeneous or conflicting.}  {The contributions are:}
\begin{enumerate}
    \item {\textit{A map of the evidence base:} a PRISMA-style accounting of 111 studies by domain and method, showing where neurophysiological evidence exists and where it does not.}
    \item {\textit{A synthesis of the musical-improvisation literature} across fMRI, EEG, fNIRS, hyperscanning, neurofeedback, and stimulation, with an explicit account of agreements and disagreements and of the role of expertise.}
    \item {\textit{An engineering outlook for musical improvisation:} wearable EEG--fNIRS sensing and closed-loop feedback, formulated as an expertise-gated control problem and discussed as a possible complement to (but not a replacement for) deliberate practice. }
    \item {\textit{A review of improvisation in other domains} (dance, theater, sport, education, therapy, professional practice), including the study designs that dominate it, its behavioral, psychophysiological, and early neural findings, and the evidence still missing.}
\end{enumerate}

{
\textbf{Positioning against prior surveys.}
While recent reviews have synthesized the neural correlates of improvisation in music \cite{Beaty2015rev, Landau2020} and divergent thinking \cite{Wu2015}, few reviews cover several domains or bridge descriptive neuroscience and prescriptive engineering (see Table~\ref{tab:survey_comparison}).
}

\begin{table*}[t]
\centering
\caption{{\textbf{Positioning against prior surveys.} Each prior review is summarized in its own terms; the final column states the advance this survey makes relative to it.}}
\label{tab:survey_comparison}
{
\renewcommand{\arraystretch}{1.25}
\footnotesize
\begin{tabular}{|p{2.4cm}|p{6.5cm}|p{6.8cm}|}
\hline
\textbf{Survey (Year)} & \textbf{What it contributed} & \textbf{Gap addressed by this survey} \\
\hline
Dietrich (2004)~\cite{dietrich2004flow} & Proposed transient hypofrontality as a mechanism of flow, linking effortless performance to temporary down-regulation of prefrontal executive control; theory-driven, with sport and free jazz as examples. & Predates the improvisation-specific imaging literature; we report how later music studies support, qualify, or contradict the prefrontal account. \\
\hline
Beaty (2015)~\cite{Beaty2015rev} & Synthesis of fMRI studies of musical improvisation (jazz, classical, freestyle rap) highlighting a prefrontal network (pre-SMA, mPFC, IFG, DLPFC, dorsal premotor cortex) and default-mode involvement. & fMRI only and no non-musical domains; we add EEG, fNIRS, hyperscanning, neurofeedback, and stimulation evidence and map non-musical domains. \\
\hline
Loui (2018)~\cite{Loui2018} & Framed musical improvisation as a multi-level system and reviewed multimodal neuroimaging of idea generation and evaluation. & Descriptive and music-only; we add an engineering outlook with an explicit control formulation. \\
\hline
Faber \& McIntosh (2020)~\cite{Faber2020} & Proposed a ``standard model'' of musical improvisation integrating creativity, production, and auditory monitoring. & Auditory-motor and correlational; we add a quantified corpus, an expertise-focused comparison of conflicting findings, and a closed-loop roadmap. \\
\hline
Goldman (2025)~\cite{Goldman2024} & Critically examined music-neuroscience methodology (mechanistic explanation, cross-domain comparison, reverse inference). & Methodological rather than synthetic; we provide the study-level synthesis and adopt its caution on reverse inference. \\
\hline
\rowcolor{blue!8}
\textbf{This work} & \multicolumn{2}{p{13.4cm}|}{A survey of 111 studies across music and other domains, with a major focus on music, where most neuroimaging evidence lies. It reports heterogeneous and conflicting evidence transparently, summarizes behavioral and early neural evidence for other domains, and outlines expertise-gated closed-loop BCIs as a possible complement to practice. The current evidence does not point to a universal neural signature of improvisation.} \\
\hline
\end{tabular}}
\end{table*}

\subsection{Methods of Survey and Scope}
\label{sec:methods}

We searched \textit{PubMed}, \textit{Web of Science}, \textit{Scopus}, and \textit{IEEE Xplore} with queries combining behavioral terms related to improvisation with methodological descriptors, e.g., (improvisation AND EEG), (creativity AND fMRI AND divergent thinking), (spontaneous performance AND neuroimaging), and (neurofeedback AND flow AND music).

{\textit{Eligibility.} Eligible studies involved human participants, were published in English, and examined improvisation or real-time creative generation. The corpus is tiered by evidence type: (i) studies with a genuine improvisational task (as opposed to pre-planned, memorized, or tightly constrained action) that report neurophysiological or causal-intervention data (EEG, fMRI, fNIRS, hyperscanning, neurofeedback, stimulation); (ii) neurophysiological studies of creative ideation and divergent thinking, retained as bridge evidence; and (iii) behavioral, qualitative, psychophysiological, and simulation studies, retained as contextual evidence, which in theater, dance, and applied settings are the only evidence available. Studies were excluded if they addressed divergent thinking without neurophysiological recording or an improvisational component, relied solely on behavioral ratings with no bearing on the generative process, or examined creative products only.}

{\textit{Selection.}} Searches in PubMed (312), Web of Science (274), Scopus (198), and IEEE Xplore (89) yielded 873 records; after removing 201 duplicates, 672 were screened by title and abstract. Of these, 389 were excluded: 187 addressed divergent thinking without neurophysiological recording or a performance component; 94 relied solely on behavioral ratings; 68 examined creative products rather than the generative process; and 40 were not peer-reviewed English-language publications. Of the 283 full texts assessed, 172 {reported neither quantifiable neurophysiological or causal-intervention data nor contextual evidence on improvisation in theater, dance, or applied settings and were excluded}. The final corpus comprises 111 studies (Music: 37, General Creativity: 41, Theater: 13, Dance: 7, Applied: 13). {Of these, 60 report neurophysiological recordings (EEG 34, fMRI 24, psychophysiology 2) and 51 are contextual (behavioral 27, qualitative 16, simulation 8; Table~\ref{tab:disproportion}). No formal risk-of-bias instrument was applied, consistent with the scoping nature of this work; instead, sample sizes, contrasts, and designs are reported study by study (Tables~\ref{tab:improv_studies_30}--\ref{tab:eeg_compare}). Following cautions on reverse inference \cite{Poldrack2006, Goldman2024}, inferences from regional activation or spectral power to psychological processes are treated as provisional.}

\subsection{Outline}
\label{sec:outline}
{\secref{sec:foundations} defines terms and summarizes creativity neuroscience (1975--2009), including quantitative studies of creative ideation, as a bridge to improvisation. \secref{sec:landscape} quantifies the evidence base and its limits. \secref{sec:music} is the core of the survey: neural signatures of musical improvisation, with a closing synthesis of agreements and disagreements. Because only music has enough evidence to define targets, \secref{sec:vision} follows with an outlook on closed-loop BCIs for musical improvisation. \secref{sec:beyond} then reviews improvisation in dance, theater, sport, education, therapy, and professional practice, and \secref{sec:conclusion} concludes.}

\begin{figure}[t]
    \centering
    \includegraphics[width=\linewidth]{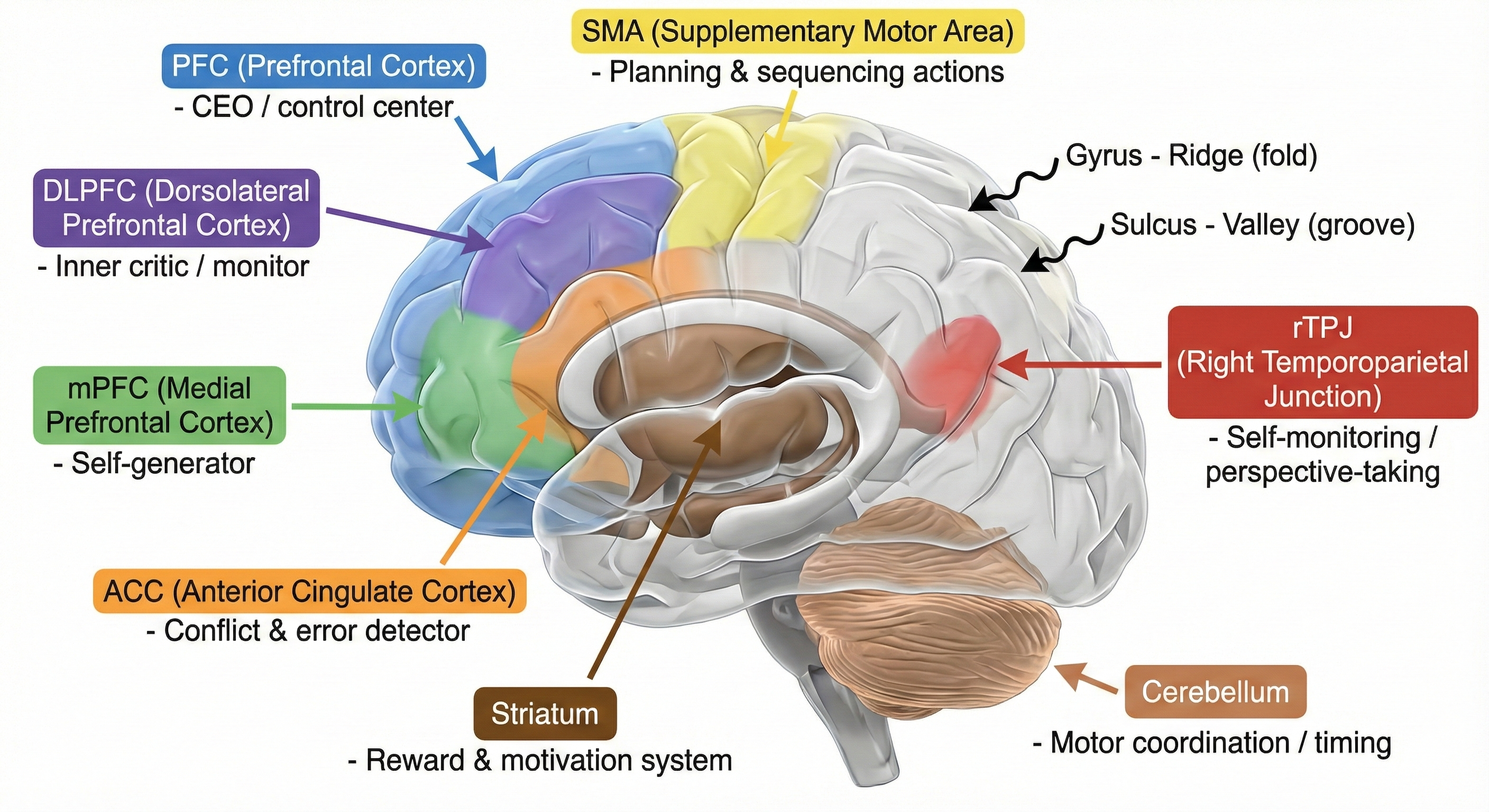}\\[2pt]
    \includegraphics[width=\linewidth]{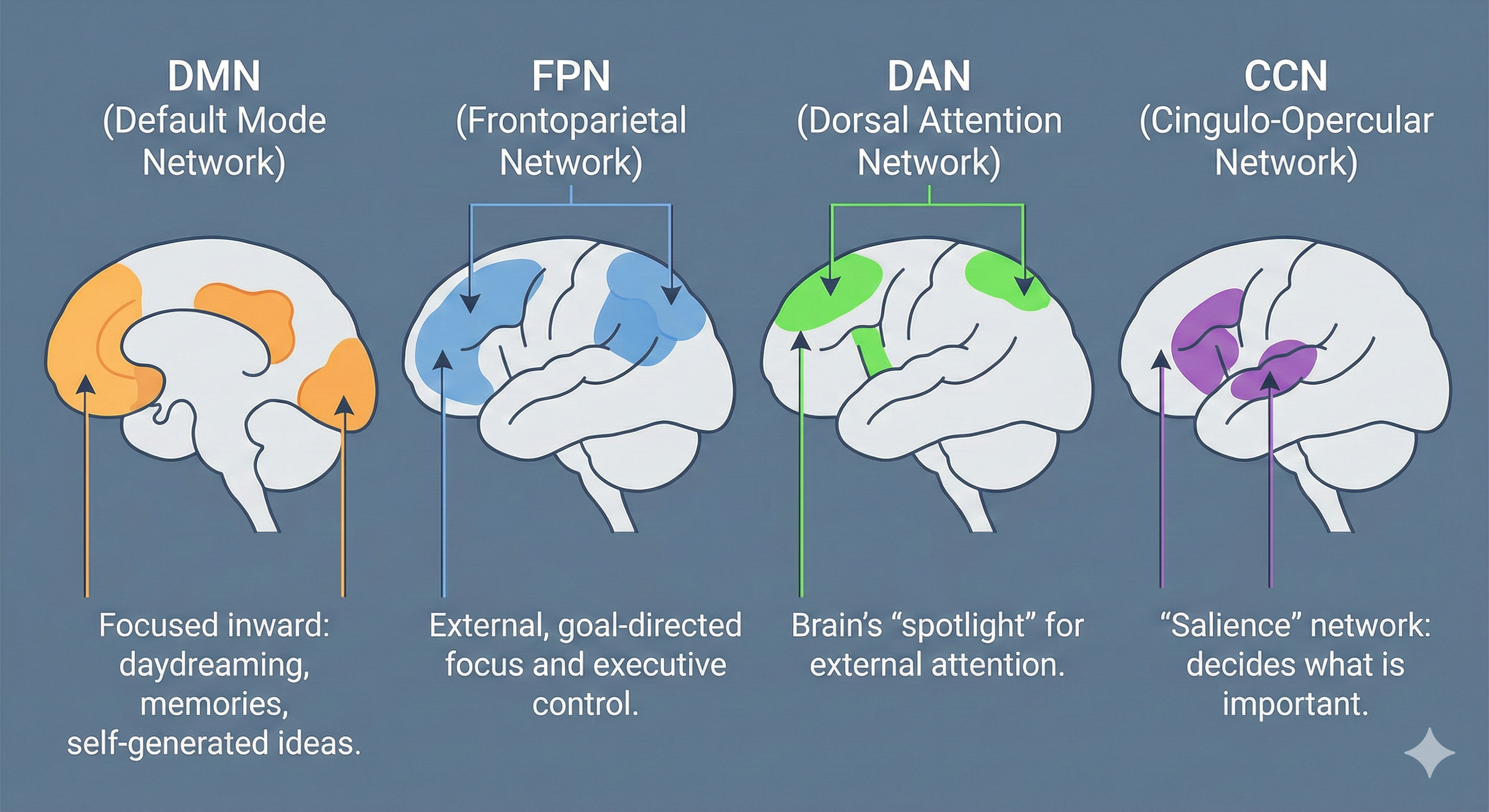}
    \caption{Brain regions (top) and networks (bottom) referred to in the text, with improvisation-related parts annotated.}
    \label{fig:anatomy}
\end{figure}

\section{Foundations: Neural Studies of Creativity as a Bridge to Improvisation}
\label{sec:foundations}

\subsection{Definitions and Measurement Primer}
{Table~\ref{tab:definitions} defines the anatomical terms, EEG bands, and concepts used in this paper, and Fig.~\ref{fig:anatomy} illustrates the regions and networks. To make later sections accessible to readers outside neuroscience, we first summarize how the main signals are measured.

\textit{fMRI} measures blood-oxygen-level-dependent (BOLD) signal changes in small volumes of brain tissue (voxels), reported in standard atlas (MNI) coordinates. Results are always relative to a comparison condition (e.g., improvisation versus playing an over-learned sequence), so ``activation'' and ``deactivation'' mean more or less BOLD signal than in that condition. \textit{EEG} records electrical activity from the scalp. Electrode positions follow the international 10--20 system: letters give the lobe (F frontal, C central, P parietal, T temporal, O occipital), odd numbers lie over the left hemisphere, even numbers over the right, and ``z'' marks the midline; Fz, Cz, and Pz are thus midline electrodes over frontal, central, and parietal cortex. EEG power is the signal energy in a frequency band (Table~\ref{tab:definitions}). \textit{fNIRS} shines near-infrared light between source and detector optodes on the scalp; each source--detector pair forms a channel over the cortex beneath it and reports changes in oxygenated hemoglobin (HbO) as a hemodynamic proxy of local activity. It is portable but reaches only the cortical surface. \textit{ERPs} (event-related potentials) are EEG responses time-locked to an event; the feedback-related negativity (FRN) reflects early evaluation of feedback, and P3a/P3b reflect later attention and updating.

For EEG studies, three ratios recur as candidate markers. The $\alpha/\theta$ ratio is alpha power divided by theta power (values above 1 mean alpha dominates). The $\beta_1/\beta_2$ ratio is low-beta (including sensorimotor rhythm) over high-beta power. \textit{Alpha/theta (A/T) neurofeedback} is not a marker but a training protocol in which trainees are rewarded for raising theta relative to alpha with eyes closed, i.e., for lowering the $\alpha/\theta$ ratio. Finally, ``bilateral'' means both hemispheres, ``lateralization'' means stronger involvement of one hemisphere, and ``medial/lateral'' and ``dorsal/ventral'' denote toward/away from the midline and upper/lower positions.}

\begin{table*}[t]
\centering
\caption{{Definitions of neuroscientific terms and EEG band taxonomy (band splits as used by Sasaki et al.\ \cite{Sasaki2019}).}}
\label{tab:definitions}
\footnotesize
{
\begin{tabular}{|p{2.3cm}|p{3.0cm}|p{2.6cm}|p{7.6cm}|}
\hline
\textbf{Category} & \textbf{Term} & \textbf{Range} & \textbf{Definition / role} \\
\hline
\multirow{10}{*}{\textbf{Regions}} 
& PFC / DLPFC & -- & Prefrontal cortex: planning, decision-making, self-regulation. DLPFC: deliberate control and monitoring (``inner critic''). \\
& mPFC & -- & Self-referential and idea-generation region. \\
& Lobes & -- & Frontal (planning, control), temporal (hearing, language, memory), parietal (spatial and sensorimotor integration), occipital (vision); orbitofrontal cortex lies on the underside of the frontal lobe (value, emotion). \\
& ACC, SMA & -- & Anterior cingulate (conflict/error detection); supplementary motor area (planning and sequencing of motor actions). \\
& Premotor, pre-SMA & -- & Motor-planning regions in front of primary motor cortex; pre-SMA initiates and sequences self-generated movement. \\
& IFG, STG & -- & Inferior frontal gyrus (sequence/syntax, speech production) and superior temporal gyrus (auditory processing); with surrounding cortex, the perisylvian language-related regions. \\
& Angular, supramarginal gyri & -- & Parietal regions for integration, attention, and language; the angular gyrus is a default-mode node. \\
& rTPJ & -- & Self-monitoring, perspective-taking, stimulus-driven attention. \\
& Cerebellum, striatum & -- & Motor timing and coordination; dopaminergic reward and motivation. \\
& DMN, FPN, DAN, CCN, MD & -- & Default-mode (internal cognition), frontoparietal (executive), dorsal attention, cingulo-opercular (salience), and multiple-demand (prefrontal--parietal, recruited by many demanding tasks) networks. \\
\hline
\multirow{9}{*}{\textbf{EEG bands}} 
& Delta ($\delta$) & 0.5--4 Hz & Deep, unconscious processing. \\
& Theta ($\theta$) & 4.5--8 Hz & Ideation, associative processing, memory. \\
& Alpha ($\alpha$) & 8.5--12 Hz & Relaxed state; cortical inhibition; internal attention. \\
& $\alpha_1$ / $\alpha_2$ & 8.5--10 / 10.5--12 Hz & External-attention inhibition / internal attention and ideation. \\
& Beta ($\beta$) & 12.5--30 Hz & Active motor planning, sensorimotor feedback, engagement. \\
& $\beta_1$ & 12.5--18 Hz & Low-level motor planning (includes SMR, 12--15 Hz, a sensorimotor idling rhythm). \\
& $\beta_2$ & 18.5--21 Hz & High-level cognitive control / executive monitoring. \\
& $\beta_3$ & 21.5--30 Hz & Motor output / muscular engagement. \\
& Gamma ($\gamma$) & 30.5--50 Hz & Information binding and perceptual integration. \\
\hline
\multirow{6}{*}{\textbf{Concepts, tools}} 
& Transient hypofrontality & -- & Temporary reduction of prefrontal executive control proposed to release spontaneous, associative processing. \\
& Alpha asymmetry & -- & Hemispheric alpha difference indexing approach/withdrawal motivation. \\
& Hyperscanning & -- & Simultaneous recording from interacting people; inter-brain synchronization. \\
& fMRI/BOLD, fNIRS & -- & Hemodynamic imaging: deep-structure (fMRI) and cortical-surface (fNIRS) oxygenation changes. \\
& Neurofeedback, BCI & -- & Real-time training via voluntary modulation of neural activity; device control or adaptation via neural signals. \\
& Reverse inference & -- & Inferring a psychological process from the activation of a region or band; provisional because regions serve many functions. \\
\hline
\end{tabular}}
\end{table*}

\subsection{Early EEG Analysis and the Alpha Signature}
Before dedicated improvisation research, EEG studies of divergent thinking linked alpha-band activity to creative ideation. Martindale and colleagues reported greater alpha activity in high scorers on creativity tasks, particularly in the inspiration stage of the creative process, and hemispheric alpha asymmetries during free fantasy speech \cite{Martindale1975, Martindale1978, Martindale1984a}---a counter-intuitive association between creativity and reduced cortical arousal. Later work added band- and region-specific detail: increased theta, delta, and beta power and greater dimensional complexity in frontal and parietal regions during divergent thinking \cite{Molle1999}; higher alpha power and distinct frontal coherence in creative than in high-IQ individuals, consistent with a \textit{neural efficiency} account \cite{jausovec2000differences, jausovec2000resting, jausovec2000complex}; and stronger frontal--posterior coupling in creative ideation than in convergent tasks, with hemispheric asymmetries \cite{Razumnikova2000, Razumnikova2004, Razumnikova2005, Razumnikova2007}. Developmental data suggest that higher-ability children recruit parietal rather than frontal resources during hypothesis generation \cite{Jin2006}, a neural-efficiency pattern that anticipates the expertise effects.

\subsection{The Hypofrontality Hypothesis}
Transient hypofrontality---a temporary down-regulation of prefrontal executive control that releases associative processing---was formalized by Dietrich \cite{Dietrich2003} and applied to flow and creative performance \cite{dietrich2004flow, dietrich2004cognitive}. Consistent with it, divergent-thinking training increased frontal alpha synchronization \cite{fink2006divergent}, and higher originality was linked to posterior and frontal alpha synchronization \cite{fink2006eeg, fink2009creative}; increases in $\beta_2$ and $\gamma$ during creative problem solving \cite{Danko2009} indicate that high-frequency binding also supports creative processing.

\subsection{Localization of Distributed Networks}
Neuro imaging further mapped where these dynamics arise. Early regional cerebral blood flow work implicated broad frontal and temporoparietal circuits \cite{Carlsson2000}. fMRI, PET, and SPECT studies (methods that map blood flow, metabolism, or tracer uptake) challenged a simple right-hemisphere account of creativity: they showed engagement of both hemispheres in prefrontal cortex and a distributed system spanning the prefrontal, orbitofrontal, temporal, parietal, and cerebellar regions \cite{Goel2005, HowardJones2005, Folley2005, ChavezEakle2007, FinkGrabner2009b}, with considerable left-hemisphere engagement during divergent tasks \cite{FinkGrabner2009a}. Related work on fluid analogies and discourse inference also showed distributed, bilateral recruitment \cite{geake2005neural, Sieborger2007}. fNIRS in musicians showed \textit{increased} prefrontal oxygenation during divergent thinking \cite{Gibson2009}.

\subsection{Beyond Functional Imaging: Structural, Chemical, and Behavioral Correlates}
Imaging research also explored the structural and neuromodulatory architecture of creativity. De Manzano et al.\ reported that divergent thinking is negatively correlated with thalamic D2 binding potentials, suggesting that reduced dopaminergic gating facilitates creative ideation \cite{demanzano2010thinking}. Jung and colleagues linked DTI and spectroscopy measures with divergent thinking, reporting inverse relationships between white matter integrity and creativity consistent with neural-efficiency models \cite{jung2010white, jung2009biochemical}, and Moore et al.\ reported negative correlations between corpus callosum volume and creativity scores \cite{Moore2009}.

Behavioral work established the utility of improvisational states: integrating improvisational tasks into primary-school arts education improves divergent thinking \cite{Sowden2015}. Creative self-efficacy predicts creative output \cite{Tierney2002} and can be strengthened by structured mastery experiences \cite{Berkemeyer2025}, in line with Bandura's framework \cite{Bandura1977, Bandura1982}. These psychological factors are not neural findings, but they are relevant to the expertise effects reviewed below.

\subsection{{Quantitative Studies of Creative Ideation: Lessons Learned}}
\label{sec:ideation}
{Table~\ref{tab:ideation} lists the quantitative ideation studies cited above. They used slow, unconstrained tasks such as divergent-thinking tests and word-association problems, and measured regional blood flow, EEG, fMRI, or fNIRS. Four messages carry over to improvisation research. (i) Alpha power tends to rise with the originality of ideas \cite{fink2009creative, Razumnikova2007}, making frontal and parietal alpha a natural candidate marker. (ii) More creative or higher-ability individuals often show less, not more, frontal activity, the neural-efficiency pattern that motivates the expertise analyses of \secref{sec:music_expertise}. (iii) Creative thinking engages a distributed system in both hemispheres \cite{Carlsson2000, ChavezEakle2007}, so an account confined to one region or hemisphere should be treated cautiously. (iv) fNIRS studies find \textit{increased} prefrontal oxygenation during ideation \cite{Folley2005, Gibson2009}, so hypofrontality cannot simply be carried over from ideation to performance. A limit applies to all four: ideation tasks are slow and unpaced, unlike real-time performance, so these studies suggest which signals to measure but do not themselves show how the brain improvises.}

\begin{table*}[ht]
\centering
\caption{{Quantitative studies of creative ideation (bridge evidence). N: number of participants; SPECT: single-photon emission computed tomography, a tracer-based imaging method. Other terms are defined in Table~\ref{tab:definitions}.}}
\label{tab:ideation}
\scriptsize
\setlength{\tabcolsep}{3pt}
\begin{tabular}{@{}p{3.5cm}p{1.3cm}p{2.5cm}p{3.6cm}p{3.0cm}p{3.1cm}@{}}
\toprule
\textbf{Study} & \textbf{N} & \textbf{Participants} & \textbf{Task} & \textbf{Measure} & \textbf{Brain focus} \\
\midrule
Carlsson et al., 2000 \cite{Carlsson2000} & 24 & Male undergraduates & Divergent-thinking tests & Regional cerebral blood flow & Prefrontal, frontotemporal \\
Ch\'avez-Eakle et al., 2007 \cite{ChavezEakle2007} & 12 & Adults & Verbal creativity test & SPECT (blood flow) & Prefrontal, cerebellum \\
Folley \& Park, 2005 \cite{Folley2005} & 51 & Adults & Divergent thinking (verbal creativity) & fNIRS & Prefrontal \\
Razumnikova, 2007 \cite{Razumnikova2007} & 39 & University students & Word-association (remote associates) tasks & EEG power and coherence & Frontal, parietal \\
Fink et al., 2009a \cite{fink2009creative} & 47 (EEG), 21 (fMRI) & University students & Divergent thinking & EEG and fMRI & Frontal--parietal \\
Gibson et al., 2009 \cite{Gibson2009} & 55 & Musicians and non-musicians & Divergent-thinking tasks & Prefrontal oxygenation (fNIRS) & Prefrontal cortex \\
\bottomrule
\end{tabular}
\end{table*}

\section{The Evidence Base: Composition and Limits}
\label{sec:landscape}

{Table~\ref{tab:disproportion} shows that the 111-study corpus is dominated by musical improvisation (37 studies; EEG 15, fMRI 14) and general creativity tasks (41; EEG 18, fMRI 10), together 70.3\% of the total. Theater contributes 13 studies (one EEG, no fMRI), dance 7 (behavioral), and applied settings 13 (simulations and qualitative methods); EEG also appears in two dance-related anchor studies (dancers' alpha activity during creative-thinking tasks \cite{Fink2009b}, and dance neurofeedback \cite{Gruzelier2014d}), which Table~\ref{tab:disproportion} counts under their primary task. Only 60 of the 111 studies contain neurophysiological recordings, and the median music study has 14 participants. Roughly half of the improvisation literature is therefore behavioral, qualitative, or simulation-based. This distribution reflects the field itself; we accordingly summarize the neural literature for music in detail and, for other domains, describe the behavioral, psychophysiological, and early neural evidence that exists. Table~\ref{tab:improv_studies_30} lists the anchor studies of improvisation: all studies that recorded neural activity during an improvisational task are musical, and the few non-musical studies with neural recordings (discussed in \secref{sec:beyond}) used other recording contexts. Quantitative ideation studies, which are bridge evidence rather than improvisation studies, were already provided in Table~\ref{tab:ideation} (\secref{sec:ideation}).}

\begin{table*}[ht]
\centering
\caption{{\textbf{Quantitative Disproportion Summary by Domain.} Distribution of the 111 reviewed studies across domains, with sample sizes, modalities, task durations, and key metrics, quantifying the auditory-motor skew of the corpus.}}
\label{tab:disproportion}
{
\begin{tabular}{|p{2.3cm}|p{1.4cm}|p{2.6cm}|p{3.4cm}|p{1.8cm}|p{1.8cm}|p{2.6cm}|}
\hline
\textbf{Domain} & \textbf{No. of Studies} & \textbf{Sample Size} & \textbf{Modalities} & \textbf{Task Duration} & \textbf{Expertise Reported} & \textbf{Key Metrics} \\
\hline
Music (Auditory) & 37 & N=3–55 (median 14) & EEG: 15, fMRI: 14, Behavioral: 5, Qualitative: 3 & 9–45 min & Yes (all) & Spectral power, BOLD, MIDI \\
\hline
General Creativity & 41 & N=12–51 (median 24) & EEG: 18, fMRI: 10, Behavioral: 11, Qualitative: 2 & 2–60 min & Partial & Alpha power, rCBF, coherence \\
\hline
Theater (Social) & 13 & N=28–64 (median 35) & Behavioral: 4, Qualitative: 6, Psychophys: 2, EEG: 1 & Variable & Partial & EDA, HR, DASS, surveys \\
\hline
Dance (Somatic) & 7 & N=32–64 (median 48) & Behavioral: 7, EEG: 0, fMRI: 0 & 12 wk–1 hr & Partial & Motion complexity, HRV \\
\hline
Applied / Org & 13 & N=1–103 (varied) & Simulations: 8, Qualitative: 5 & Variable & Rarely & Org. outcomes, error rates \\
\hline
\end{tabular}
}
\end{table*}

\begin{table*}[ht]
\centering
\caption{{Anchor studies of improvisation: sample, task, measure, and brain focus. The first group contains music studies. The second group contains the few non-musical studies with neural recordings; these were obtained during creative-thinking tasks, imagery, neurofeedback, or before and after improvisation training, not during improvised performance. Quantitative creative-ideation studies are listed separately in Table~\ref{tab:ideation}. N: number of participants; NF: neurofeedback; SMR: sensorimotor rhythm; HRV: heart-rate variability. Other terms are defined in Table~\ref{tab:definitions}.}}
\label{tab:improv_studies_30}
\scriptsize
\setlength{\tabcolsep}{3pt}
\begin{tabular}{@{}p{3.5cm}p{1.3cm}p{2.5cm}p{3.6cm}p{3.0cm}p{3.1cm}@{}}
\toprule
\textbf{Study} & \textbf{N} & \textbf{Participants} & \textbf{Task} & \textbf{Measure} & \textbf{Brain focus} \\
\midrule
\multicolumn{6}{@{}l}{\textit{Music studies}} \\
Limb \& Braun, 2008 \cite{Limb2008} & 6 & Jazz pianists & Jazz improvisation vs.\ over-learned playing & fMRI (BOLD signal) & Prefrontal, motor, limbic \\
Liu et al., 2012 \cite{liu2012neural} & 12 & Freestyle rappers & Lyrical improvisation vs.\ memorized lyrics & fMRI & Medial and dorsolateral prefrontal \\
Berkowitz \& Ansari, 2008 \cite{berkowitz2008generation} & 12 & Pianists & Melodic and rhythmic improvisation & fMRI & Premotor, prefrontal, cingulate \\
Berkowitz \& Ansari, 2010 \cite{berkowitz2010expertise} & 24 & Musicians and non-musicians & Melodic improvisation & fMRI & Right temporoparietal junction \\
de Manzano \& Ull\'en, 2012 \cite{deManzano2012a} & 15 & Concert pianists & Free melodic and rhythmic improvisation & fMRI & Pre-SMA, premotor, inferior frontal \\
Donnay et al., 2014 \cite{Donnay2014} & 11 & Jazz pianists & Interactive improvisation (trading fours) & fMRI & Language-related and default-mode regions \\
Pinho et al., 2014 \cite{pinho2014connecting} & 39 & Classical and jazz pianists & Improvisation, by years of experience & fMRI activity and connectivity & Frontoparietal executive, DLPFC, premotor \\
Lu et al., 2017 \cite{LuYao2017} & 29 & Composers & Melody improvisation with cues & fMRI & Multiple-demand system \\
Sasaki et al., 2019 \cite{Sasaki2019} & 14 & Guitarists & Improvisation vs.\ scale playing & EEG power & Prefrontal, supplementary motor, temporal \\
Rosen et al., 2020 \cite{Rosen2020} & 32 & Jazz guitarists & Improvisations rated for quality & EEG power & Frontal, central, parietal \\
Lopata et al., 2017 \cite{Lopata2017} & 22 & Musicians & Improvisation vs.\ listening & EEG alpha power & Frontal upper alpha \\
Yaghmour et al., 2021 \cite{Yaghmour2021} & 1 & Ney player & Maqam improvisation & EEG power spectra & Left frontal and temporal \\
Faller et al., 2021 \cite{Faller2021} & 40 & Musicians & Listening to chords with unexpected events & EEG connectivity and power & Default-mode, salience, attention, executive networks \\
Ramirez-Moreno et al., 2023 \cite{RamirezMoreno2023} & 3 & Musicians & Live jazz performance & EEG inter-brain synchrony & Temporal--occipital \\
Maidhof et al., 2023 \cite{Maidhof2023} & 2 & Therapist and client & Drum improvisation in music therapy & EEG inter-brain coupling & Frontal, central \\
Gruzelier et al., 2014 (immediate) \cite{Gruzelier2014} & 24 & Music students & Instrumental or vocal improvisation after NF & EEG neurofeedback; ratings & Alpha/theta at Pz; SMR at Cz \\
Gruzelier et al., 2014 (replication) \cite{Gruzelier2014b} & 19 & Music students & Singing or instrumental improvisation after NF & EEG neurofeedback; ratings & Alpha, theta, SMR, high beta \\
Gruzelier et al., 2014 (children) \cite{Gruzelier2014c} & 30 & 11-year-olds & Improvisation and attention test after NF & EEG neurofeedback; errors & SMR at Cz; theta/alpha at Pz \\
\midrule
\multicolumn{6}{@{}l}{\textit{Other domains}} \\
Fink et al., 2009b \cite{Fink2009b} & 32 & Professional dancers and novices & Creative thinking; imagery of dance improvisation & EEG alpha power change & Parietal, right hemisphere \\
Gruzelier et al., 2014 (dancers) \cite{Gruzelier2014d} & 64 & Dance students & Alpha/theta neurofeedback and HRV training & EEG; heart-rate variability & Alpha/theta at Pz; creativity, anxiety \\
DeMichele \& Kuenneke, 2021 \cite{DeMichele2021} & 28 & Adolescents & Short-form improv training, before and after & Quantitative EEG (qEEG) & Whole-brain coherence \\
Sepp\"anen et al., 2020 \cite{Seppanen2020} & 35 & Student teachers & Social stress test after improv training & EEG, heart rate, skin conductance & Physiological stress responses \\
Rosso et al., 2025 (unpublished) \cite{Rosso2025} & 32 & Healthy adults & Flow and attention tasks & EEG power, alpha asymmetry & Medial and dorsolateral prefrontal \\
\bottomrule
\end{tabular}
\end{table*}

\subsection{{Limitations of the Evidence Base}}
{Four limitations of the published literature constrain inference. (1) \textit{Ecological validity:} improvisation is typically studied under constraints (immobility in the scanner, scales or short pieces), unlike real performance or real-world improvisation \cite{Limb2008, Lopata2017}. (2) \textit{Heterogeneous contrasts:} improvisation is contrasted with scale playing, memorized or over-learned playing, listening, or expert quality ratings, which dissociate different processes. (3) \textit{Statistical power and reporting:} samples are small and effect sizes and control conditions are inconsistently reported. (4) \textit{Single-brain focus:} few adequately powered studies address inter-brain dynamics, and simultaneous EEG--fNIRS recording during performance is essentially absent.}

\section{Neural Signatures of Musical Improvisation}
\label{sec:music}

{Musical improvisation is the only domain with a neuroimaging literature large enough to compare studies (Table~\ref{tab:disproportion}). This section reviews it by modality (fMRI, EEG), then by network-level, expertise, social, and causal evidence, and closes with an explicit account of where studies agree and disagree. Each finding is reported with its sample, contrast, and modality so that readers can judge its generality.}

\subsection{Hemodynamic Evidence: fMRI}
\label{sec:music_fmri}
{The best-known fMRI result comes from jazz pianists: improvisation, compared with over-learned sequences, was characterized by extensive deactivation of dorsolateral prefrontal and lateral orbital regions together with focal activation of medial prefrontal (frontal polar) cortex, accompanied by widespread sensorimotor activation and limbic deactivation \cite{Limb2008}. A similar dissociation (DLPFC down, mPFC up) was reported for freestyle rap relative to memorized lyrics \cite{liu2012neural}. Other fMRI studies emphasize different regions: melodic and rhythmic improvisation in pianists modulated a network of dorsal premotor cortex, rostral cingulate zone (part of the anterior cingulate cortex), and inferior frontal gyrus \cite{berkowitz2008generation}; free melodic and rhythmic improvisation engaged pre-SMA and dorsal premotor areas \cite{deManzano2012a}; musicians (but not non-musicians) deactivated rTPJ during melodic improvisation \cite{berkowitz2010expertise}; and trading fours in jazz activated inferior frontal and posterior superior temporal (perisylvian, language-related) areas while deactivating angular and supramarginal gyri \cite{Donnay2014}. In composers improvising melodies, the multiple-demand system (a prefrontal--parietal network recruited by many demanding tasks) was significantly activated, with parietal involvement increasing for unfamiliar cues \cite{LuYao2017}. Beaty's review of the fMRI literature lists the pre-SMA, mPFC, IFG, DLPFC, and dorsal premotor cortex as a commonly engaged prefrontal network and notes that premotor and lateral prefrontal activation suggests that seemingly unconstrained behavior may benefit from motor planning and cognitive control, while midline activation points to spontaneous, default-network cognition \cite{Beaty2015rev}. Table~\ref{tab:fmri_compare} summarizes the studies.}

\begin{table*}[ht]
\centering
\caption{{\textbf{fMRI studies of musical improvisation.} Sample, contrast, and principal findings as reported by the original studies.}}
\label{tab:fmri_compare}
{
\footnotesize
\renewcommand{\arraystretch}{1.12}
\begin{tabular}{|p{3.1cm}|p{3.0cm}|p{3.8cm}|p{5.8cm}|}
\hline
\textbf{Study} & \textbf{Sample} & \textbf{Contrast} & \textbf{Main finding} \\
\hline
Limb \& Braun, 2008 \cite{Limb2008} & 6 jazz pianists & Improvisation vs.\ over-learned playing & DLPFC and lateral orbital deactivation; focal mPFC activation \\
\hline
Liu et al., 2012 \cite{liu2012neural} & 12 freestyle rappers & Improvised vs.\ memorized lyrics & DLPFC down, mPFC up \\
\hline
Berkowitz \& Ansari, 2008 \cite{berkowitz2008generation} & Trained pianists & Melodic and rhythmic improvisation vs.\ controls & Dorsal premotor cortex, rostral cingulate zone, inferior frontal gyrus \\
\hline
Berkowitz \& Ansari, 2010 \cite{berkowitz2010expertise} & Musicians vs.\ non-musicians & Melodic improvisation & rTPJ deactivation in musicians only \\
\hline
de Manzano \& Ull\'en, 2012 \cite{deManzano2012a} & 15 concert pianists & Free melodic $\times$ rhythmic improvisation & Pre-SMA and dorsal premotor activation and connectivity \\
\hline
Donnay et al., 2014 \cite{Donnay2014} & 11 jazz pianists (pairs) & Interactive improvisation (trading fours) & IFG and posterior STG activation; angular and supramarginal deactivation \\
\hline
Pinho et al., 2014 \cite{pinho2014connecting} & 39 pianists (classical and jazz) & Improvisation; hours of improvisation experience & Experience negatively associated with frontoparietal executive activity; positively with connectivity of bilateral DLPFC, dorsal premotor, and pre-SMA \\
\hline
Lu et al., 2017 \cite{LuYao2017} & 29 composers & Melody improvisation with familiar vs.\ unfamiliar cues & Multiple-demand system activated; parietal (IPS) increase for unfamiliar cues \\
\hline
\end{tabular}}
\end{table*}

\subsection{Oscillatory Evidence: EEG}
\label{sec:music_eeg}
Temporal dynamics refer to sub-second to seconds-scale fluctuations in brain activity that fMRI averages over. {Source-resolved EEG of 14 skilled guitarists showed greater theta, alpha, and beta power for improvisation than for scale playing in medial frontal, middle frontal, anterior cingulate, polar medial prefrontal, premotor, and temporoparietal sources, and single-trial classification of improvisation versus scale playing exceeded 75\% \cite{Sasaki2019}. Frontal upper-alpha power rose during improvisation, most clearly in musicians with improvisation training \cite{Lopata2017}. A single-participant study of ney improvisation reported low-frequency (theta, alpha) increases over left frontal and temporal areas and high-frequency (high-beta, gamma) increases over right temporal and left parietal areas \cite{Yaghmour2021}. Relating EEG to expert ratings in 32 jazz guitarists, higher-quality improvisations were associated with beta/gamma activity in predominantly posterior left-hemisphere clusters and lower-quality ones with right temporoparietal and fronto-polar activity; after controlling for experience, quality tracked right-frontal activity \cite{Rosen2020} (see Table~\ref{tab:eeg_compare}).

Two points follow. First, the sub-band taxonomy of Table~\ref{tab:definitions} is applied in full by only one study; the others report broad-band alpha or beta, so sub-band comparison across studies is not yet possible, and the direction of effects depends on the contrast (versus scale playing, listening, or quality ratings). Second, only two studies resolve time-courses: in a turn-taking piano duet, the feedback-related negativity to altered feedback was smaller during improvised than score-played turns, whereas the later P3a/P3b did not differ, indicating reduced early performance monitoring when improvising \cite{Kim2025}; and in live jazz, beta/gamma inter-brain synchronization rose with the number of musicians playing together, with positive bispectrum changes about three seconds before synchronized events \cite{RamirezMoreno2023}. No study reports onset and decay of a spectral marker around the transition into improvisation; this is the key missing quantity for closed-loop state detection (this will be discussed more in \secref{sec:control_formulation}).}
\begin{table*}[ht]
\centering
\caption{{\textbf{EEG studies of musical improvisation: sample, contrast, and findings.} Entries restate the primary reported results; ``$\uparrow$'' denotes increased power or amplitude for the first-named condition.}}
\label{tab:eeg_compare}
{
\footnotesize
\renewcommand{\arraystretch}{1.15}
\begin{tabular}{|p{2.5cm}|p{2.9cm}|p{3.3cm}|p{2.2cm}|p{5.2cm}|}
\hline
\textbf{Study} & \textbf{Sample} & \textbf{Contrast} & \textbf{Measure} & \textbf{Main finding} \\
\hline
Lopata et al., 2017 \cite{Lopata2017} & 22 musicians; improvisation-trained vs.\ not & Improvisation vs.\ other conditions & Frontal upper alpha & Alpha $\uparrow$ during improvisation, strongest with improvisation training \\
\hline
Sasaki et al., 2019 \cite{Sasaki2019} & 14 skilled guitarists & Improvisation vs.\ scale (32-s blocks) & $\theta$, $\alpha$, $\beta$; sources & Power $\uparrow$ in MFC, MFG, ACC, polar mPFC, PMC, STG, IPL, TPJ; $>$75\% single-trial classification \\
\hline
Rosen et al., 2020 \cite{Rosen2020} & 32 jazz guitarists; graded experience & High- vs.\ low-rated improvisations & $\beta$, $\gamma$ clusters & High-rated: posterior left; low-rated: right TPJ and fronto-polar; right-frontal after controlling for experience \\
\hline
Yaghmour et al., 2021 \cite{Yaghmour2021} & 1 professional ney player & Maqam improvisation & $\theta$, $\alpha$; high-$\beta$, $\gamma$ & Low bands $\uparrow$ left frontal/temporal; high bands $\uparrow$ right temporal/left parietal \\
\hline
Kim et al., 2025 \cite{Kim2025} & Piano duets & Improvised vs.\ score turns; altered feedback & FRN, P3a/P3b (ERP) & FRN smaller when improvising; P3a/P3b not different (early, not late, monitoring reduced) \\
\hline
Ramirez-Moreno et al., 2023 \cite{RamirezMoreno2023} & 3 musicians, 45-min live jazz & Together vs.\ solo; synchronous vs.\ not & Inter-brain bispectrum, 13--50 Hz & Beta/gamma synchrony $\uparrow$; changes $\approx$3 s before synchronized events \\
\hline
\end{tabular}}
\end{table*}

\subsection{Network-Level Evidence and Flow}
\label{sec:music_networks}
Faller et al.\ identified spatiospectral EEG networks, including default-mode, cingulo-opercular, dorsal-attention, and frontoparietal components, whose expression varied with improvisational experience \cite{Faller2021}. In fMRI, improvisation experience was associated with stronger functional connectivity among bilateral DLPFC, dorsal premotor, and pre-SMA regions \cite{pinho2014connecting}. The default-mode network's role in creative cognition is reviewed in \cite{Beaty2019}. Flow, the psychophysiological correlate of fluent performance \cite{csikszentmihalyi1990flow}, has been linked to prefrontal EEG markers in a preprint \cite{Rosso2025}.

\subsection{Expertise}
\label{sec:music_expertise}
{Expertise is the most consistent modulator across modalities, although the evidence comes from few laboratories. In 39 pianists, hours of improvisation experience were negatively associated with frontoparietal executive activity, which the authors interpret as automation of a creative behavior \cite{pinho2014connecting}. Musicians, but not non-musicians, deactivated rTPJ during melodic improvisation \cite{berkowitz2010expertise}. Causally, anodal tDCS over right DLPFC produced no main effect on rated improvisation quality but a significant stimulation$\times$expertise interaction: it helped less experienced and hindered more experienced jazz pianists \cite{Rosen2016}. In EEG, the balance between executive and associative activity depended on experience \cite{Rosen2020}. Two caveats apply: several widely cited studies include only experts (\cite{Limb2008, liu2012neural, Sasaki2019}) and therefore cannot show an expert--novice difference, and most of the expertise evidence comes from jazz paradigms.}

\subsection{Social and Interactive Improvisation}
Improvisation is inherently interactive \cite{Nelson2024}. EEG hyperscanning shows inter-brain synchronization increasing with coordinated musical performance \cite{RamirezMoreno2023, Maidhof2023, Walton2018, Kim2025}; in interactive jazz, perisylvian language-related areas were engaged \cite{Donnay2014}. All hyperscanning studies in the corpus are single-case or very small samples.

\subsection{Causal Evidence: Neurofeedback and Stimulation}
\label{sec:music_causal}
{Besides the tDCS result above \cite{Rosen2016}, neurofeedback offers the strongest causal evidence. In the alpha/theta (A/T) protocol, performers learn, eyes closed, to raise theta over alpha at Pz; in the sensorimotor-rhythm (SMR) protocol they raise SMR over theta and high beta. In conservatoire musicians, A/T training improved the creativity/musicality, technique, and communication of advanced playing and extended to novice prepared and improvised performance, whereas SMR training benefited novice performance only, consistent with its effect on lower-order attention and working-memory processes \cite{Gruzelier2014, Gruzelier2014b}. In 11-year-olds, A/T and SMR training improved creative improvisation \cite{Gruzelier2014c}. Two caveats apply: these trials improved rated performance after training rather than targeting the state during performance, and the A/T protocol trains theta-over-alpha at rest, the opposite spectral direction to the in-performance alpha increases seen in EEG studies of improvisation.}

\subsection{{Synthesis: Agreement and Disagreement}}
\label{sec:music_synthesis}
{Table~\ref{tab:disagree} gathers the main points on which the music literature agrees or differs. Three observations follow.

\textit{The prefrontal picture is not uniform.} Dissociated prefrontal activity (DLPFC down, mPFC up) is reported in two fMRI studies with expert samples and an over-learned or memorized comparison condition \cite{Limb2008, liu2012neural}. Other fMRI work and reviews report lateral prefrontal and premotor engagement \cite{Beaty2015rev, LuYao2017}, EEG reports increased prefrontal power in experts \cite{Sasaki2019}, and fNIRS in musicians shows increased prefrontal oxygenation during divergent thinking \cite{Gibson2009}. These findings are not necessarily incompatible: they differ in contrast condition, task (performance versus ideation), modality (BOLD, oxygenation, spectral power), and expertise, and spectral power cannot be mapped onto BOLD deactivation without reverse inference \cite{Poldrack2006}. The literature does not yet allow the conclusion that hypofrontality is a general feature of musical improvisation, nor that it is absent.

\textit{Lateralization is unsettled.} The older notion that creativity is a right-hemisphere function is not supported as a simple account: bilateral prefrontal engagement and considerable left-hemisphere involvement were reported in creativity studies \cite{Goel2005, HowardJones2005, FinkGrabner2009a}, whereas improvisation studies report right-lateralized effects (right-frontal activity related to quality after controlling for experience \cite{Rosen2020}; right-DLPFC stimulation effects \cite{Rosen2016}; rTPJ deactivation \cite{berkowitz2010expertise}) alongside left-lateralized ones (left frontal and temporal low-frequency increases in a single ney player \cite{Yaghmour2021}; posterior-left high-frequency activity for high-rated improvisations \cite{Rosen2020}).

\textit{Expertise matters, but its form is unsettled.} Experience reduces frontoparietal activity and strengthens connectivity \cite{pinho2014connecting}, and stimulation effects reverse with expertise \cite{Rosen2016}; whether this amounts to an expert--novice \textit{inversion} of DLPFC activity remains open, because no study in the corpus compares novice and expert DLPFC activity under the same improvisation versus constrained-playing contrast.}

\begin{table*}[ht]
\centering
\caption{{\textbf{Points of agreement and disagreement in the musical-improvisation literature.}}}
\label{tab:disagree}
{
\footnotesize
\renewcommand{\arraystretch}{1.12}
\begin{tabular}{|p{2.5cm}|p{4.4cm}|p{4.6cm}|p{4.2cm}|}
\hline
\textbf{Topic} & \textbf{Reported pattern A} & \textbf{Reported pattern B} & \textbf{Likely sources of difference} \\
\hline
Lateral prefrontal (DLPFC) & Deactivation vs.\ over-learned/memorized playing (fMRI) \cite{Limb2008, liu2012neural} & Engagement of lateral prefrontal and multiple-demand regions \cite{Beaty2015rev, LuYao2017}; increased prefrontal power in EEG \cite{Sasaki2019}; increased oxygenation in fNIRS (divergent thinking) \cite{Gibson2009} & Contrast condition, task, modality, expertise \\
\hline
Medial prefrontal / default mode & mPFC activation \cite{Limb2008, liu2012neural}; midline default-network activation \cite{Beaty2015rev} & Angular-gyrus deactivation in interactive improvisation \cite{Donnay2014} & Solo vs.\ interactive task \\
\hline
Alpha/theta power & Increased frontal alpha/theta during improvisation \cite{Lopata2017, Sasaki2019, Yaghmour2021} & Concurrent beta (and gamma) increases \cite{Sasaki2019, Rosen2020}; A/T neurofeedback trains theta over alpha at rest \cite{Gruzelier2014b} & Contrast, band definitions, eyes open vs.\ closed; interpretation of alpha is inferential \\
\hline
Lateralization & Right-lateralized effects \cite{Rosen2020, Rosen2016, berkowitz2010expertise} & Bilateral/left effects \cite{Goel2005, FinkGrabner2009a, Yaghmour2021, Rosen2020} & Quality-based vs.\ condition-based contrasts; expertise; band \\
\hline
Expertise & Experience reduces frontoparietal activity, increases connectivity \cite{pinho2014connecting}; tDCS effect reverses \cite{Rosen2016} & Several key studies include experts only \cite{Limb2008, liu2012neural, Sasaki2019} & Few laboratories; jazz paradigms \\
\hline
\end{tabular}}
\end{table*}

{The next section uses these findings, with their disagreements, to outline sensing and closed-loop feedback for musical improvisation; \secref{sec:beyond} then turns to other domains.}

\section{Outlook: Closed-Loop Brain--Computer Interfaces for Musical Improvisation}
\label{sec:vision}

{This section looks forward from \secref{sec:music}. It is specified for \textit{musical} improvisation only, for the following reason. Though the sensing hardware and the control architecture of the proposed BCI framework are domain-agnostic, but a closed-loop system also needs a \textit{target state} and an \textit{expertise gate} (see Fig.~\ref{fig:bci_pipeline}). These can be defined only from neural evidence recorded during improvisation, which exists for expert musicians and, as \secref{sec:beyond} shows, essentially not for other domains. Extending the system to dance, theater, or professional practice would reuse the architecture but first require the missing neuro imaging evidence listed at the end of \secref{sec:beyond}.

Deliberate practice remains the established route to improvisational expertise, and the literature reviewed here contains no evidence that neural feedback accelerates it beyond what practice achieves. We therefore frame wearable sensing and closed-loop feedback as (i) a research tool for measuring performers' states in naturalistic settings and (ii) a possible adjunct to practice whose added value is an empirical question; the trial of \secref{sec:trial} includes practice-only and sham-feedback arms to test exactly this. Because the literature disagrees on several points (Table~\ref{tab:disagree}), the target states below are design hypotheses to be calibrated, and the controller is specified so that targets can be replaced as evidence changes.}

\subsection{{Problem Formulation: Improvisation as a Feedback Control Task}}
\label{sec:control_formulation}

{We cast maintenance of a target performance state in expert musicians as a closed-loop control problem. Let the performer's state be summarized by the feature vector
\begin{equation}
\mathbf{x}(t)=\begin{bmatrix} r_{\alpha\theta}(t)\\ \Delta h(t)\\ \rho(t)\end{bmatrix}\in\mathbb{R}^3,
\label{eq:state}
\end{equation}
where $r_{\alpha\theta}$ is the $\alpha/\theta$ ratio (alpha, 8.5--12 Hz, over theta, 4.5--8 Hz, power) at Pz in a sliding window, $\Delta h$ is the change in DLPFC oxygenated hemoglobin (HbO) from fNIRS, and $\rho=\beta_1/\beta_2$ is the low-to-high beta ratio at Cz (a sensorimotor-gating index; Table~\ref{tab:definitions} and \secref{sec:foundations} define the electrodes and bands). All features are referenced to the same performer's constrained-playing (e.g., scale-playing) baseline in the same session: $r_{\alpha\theta}$ and $\rho$ are divided by their baseline values (so 1 means no change) and $\Delta h$ is the difference from baseline. The target set depends on the expertise estimate $\hat{E}\in\{\text{novice},\text{expert}\}$, obtained from a behavioral skill set (to be discussed in \secref{sec:trial}), not from $\mathbf{x}$:
\begin{equation}
\mathcal{T}_{\text{expert}}=\{\mathbf{x}: r_{\alpha\theta}>\tau_r\ \wedge\ \Delta h\le\tau_h\},\qquad
\mathcal{T}_{\text{novice}}=\{\mathbf{x}: \rho>\tau_\rho\},
\label{eq:target}
\end{equation}
with thresholds $\tau_r=1+\delta_r$, $\tau_h=-\delta_h$, $\tau_\rho=1+\delta_\rho$ and margins $\delta\ge 0$ calibrated per performer (see below). Define the signed deviation $\mathbf{e}_{\hat{E}}(t)$ as $[\tau_r-r_{\alpha\theta},\ \Delta h-\tau_h]^{\top}$ for experts and $\tau_\rho-\rho$ for novices, so that $\mathbf{x}\in\mathcal{T}_{\hat{E}}\iff\mathbf{e}_{\hat{E}}\le\mathbf{0}$ elementwise. The controller delivers a graded, sub-disruptive cue (haptic or auditory, or relaxation of task constraints) only when the state has stayed outside the target for a dwell time $T_p$:
\begin{equation}
\mathbf{u}(t)=\begin{cases}\mathbf{0}, & \mathbf{x}(t)\in\mathcal{T}_{\hat{E}}\ \text{or}\ \text{dwell}<T_p,\\[2pt] g_{\hat{E}}\big(\mathbf{e}_{\hat{E}}(t)\big), & \text{otherwise,}\end{cases}
\label{eq:control_law}
\end{equation}
where $g_{\hat{E}}(\cdot)$ is a monotone gain. The dwell time prevents chattering and absorbs the multi-second hemodynamic lag of fNIRS relative to EEG; the hemodynamic criterion is evaluated on a trailing window while the spectral criterion updates at sub-second rate. The sign of the desired correction inverts between novices (toward greater beta/sensorimotor control, Phase~1) and experts (toward reduced frontal control, Phase~2), which is why the policy is gated by $\hat{E}$.

\textit{Choice of thresholds.} No study in the corpus reports a threshold for any improvisatory state: spectral results are group-level power contrasts, fMRI results are $t$-maps with MNI coordinates, and none reports a DLPFC--mPFC correlation coefficient. Absolute values (e.g., a fixed $\alpha/\theta$ of 1.0 or $\beta_1/\beta_2$ of 1.2) cannot be justified from current data: at an eyes-open parietal electrode alpha typically exceeds theta already, so $\alpha/\theta>1$ would hardly discriminate, and a fixed beta ratio has no empirical anchor. We therefore use only what the literature does support, namely the \textit{direction} of reported effects relative to constrained playing: increased alpha/theta power \cite{Lopata2017, Sasaki2019}, lower DLPFC activity in two expert fMRI studies \cite{Limb2008, liu2012neural}, and benefit of low-beta/SMR-type training in novices \cite{Gruzelier2014}. In baseline-normalized units these are sign criteria: $\tau_r=\tau_\rho=1$ and $\tau_h=0$ mean ``no change from the performer's own constrained playing'', and $\delta=0$ is the minimal criterion. Because a zero margin would make the controller react to noise, the margins $\delta_r,\delta_h,\delta_\rho$ are set in a calibration session (for example, so that constrained-playing windows fall in the target at a chosen low rate, in the spirit of adaptive reward thresholds in neurofeedback). The sign criteria are defensible as a minimal design hypothesis; the margins are a calibration problem. A pre-registered calibration study should estimate, within performers, the distributions of $r_{\alpha\theta}$ and $\Delta h$ in improvisation versus constrained playing, and the DLPFC--mPFC hemodynamic relation.

Three limits of this formulation should be stated. First, it is a state-feedback cueing policy, not an identified dynamical model: the plant (how a cue changes $\mathbf{x}$) is unknown and must be estimated empirically. Second, cues may themselves recruit executive monitoring, which would move $\Delta h$ in the wrong direction; sham-cue controls and low-salience (ambient, haptic) cues are therefore essential. Third, the margins $\delta$ are design parameters, not literature-derived constants.}

\subsection{{Bridging the Ecological Validity Gap: Wearable Improv Hyperscanning}}
\label{sec:instrumentation}

To validate the neural mechanisms of improvisation in high-stakes, real-world settings such as surgery and emergency response, research must move from static laboratory paradigms to mobile, multimodal systems. {We propose a hybrid EEG--fNIRS standard: active-electrode EEG with at least 32 channels sampled at 250 Hz or higher, a prefrontal fNIRS array covering DLPFC and mPFC, an inertial measurement unit (IMU) co-located with the headset, and stream synchronization at millisecond level using LabStreamingLayer so that neural events align with behavioral and environmental triggers.

Motion-artifact handling requires three layers. \textit{Hardware:} cable sway and electrode motion dominate movement artifacts in mobile EEG, so active-electrode caps with managed cabling and rigid, stable optode--scalp coupling are preferred \cite{Mullen2015}. \textit{Online processing:} artifact subspace reconstruction (ASR) removes high-amplitude transients in real time, with the burst-rejection cutoff chosen conservatively \cite{Mullen2015, Chang2020}; ICA with an IMU reference can separate movement-correlated components, and for fNIRS, motion correction (wavelet filtering performed best on real cognitive data \cite{Brigadoi2014}) and short-separation channels where hardware allows. \textit{Feature robustness:} fNIRS is less sensitive to electrical movement artifacts than EEG but remains susceptible to optode--scalp displacement \cite{Brigadoi2014}; for this reason the cue is suppressed (state held) only when the EEG criterion \textit{and} the fNIRS criterion agree (Eq.~\ref{eq:control_law}), so that a motion-induced spurious alpha increase alone cannot signal a relaxed state. Because source localization with 32 dry channels is coarse (centimeter-scale at best, and poorer for medial sources), the real-time EEG feature is sensor-level (Pz, Cz) and the DLPFC criterion rests on fNIRS; source imaging is used offline for calibration.

Published recovery rates for ambulatory pipelines vary with motion intensity; we therefore propose acceptance tests to be reported before any closed-loop deployment: (a) recovery of alpha power and single-trial improvisation-versus-constrained classification accuracy across standing, walking, and playing conditions, benchmarked against the $>$75\% accuracy reported in stationary recordings \cite{Sasaki2019}; (b) false-cue rate per minute with sham performers and with mechanically induced motion; and (c) end-to-end latency from state change to cue.}

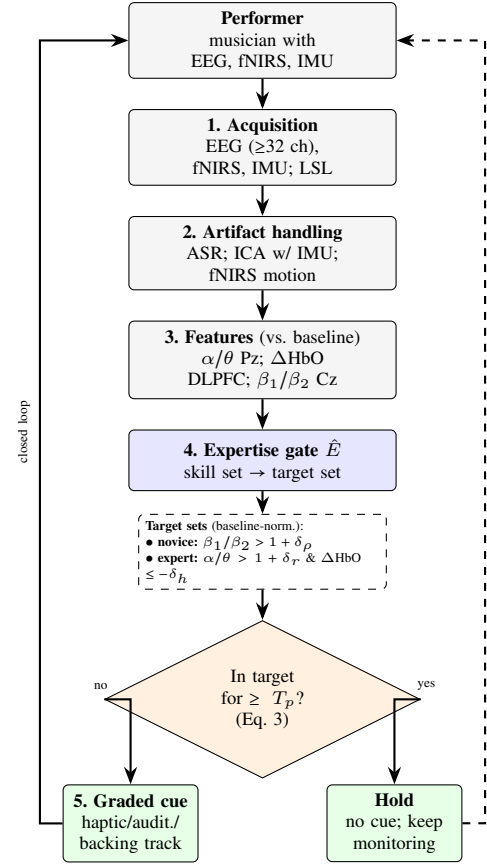
\begin{figure}[htbp]
\centering
\usetikzlibrary{positioning,arrows.meta,shapes.geometric,calc}
\begin{tikzpicture}[
  >=Stealth, font=\scriptsize,
  box/.style={draw, rounded corners=2pt, align=center, minimum height=8mm, text width=3.3cm, fill=gray!8},
  gate/.style={box, fill=blue!10},
  dec/.style={draw, diamond, aspect=2.0, align=center, inner sep=1pt, fill=orange!12, text width=2.4cm},
  outbox/.style={box, fill=green!10, text width=1.55cm, minimum height=10mm},
  arr/.style={->, thick}
]

% Main Top-to-Bottom Flow
\node[box] (perf) {\textbf{Performer}\\ musician with EEG, fNIRS, IMU};

\node[box, below=4mm of perf] (acq) {\textbf{1. Acquisition}\\ EEG ($\ge$32 ch), fNIRS, IMU; LSL};

\node[box, below=4mm of acq] (pre) {\textbf{2. Artifact handling}\\ ASR; ICA w/ IMU; fNIRS motion};

\node[box, below=4mm of pre] (feat) {\textbf{3. Features} (vs.\ baseline)\\ $\alpha/\theta$ Pz; $\Delta$HbO DLPFC; $\beta_1/\beta_2$ Cz};

\node[gate, below=4mm of feat] (gate) {\textbf{4. Expertise gate} $\hat E$\\ skill set $\rightarrow$ target set};

% Inline Target-Set Note (Embedded in main column)
\node[align=left, text width=3.1cm, below=3mm of gate, draw, dashed, rounded corners=2pt, inner sep=2.5pt, font=\tiny, fill=white] (note)
 {\textbf{Target sets} (baseline-norm.):\\ 
  $\bullet$ \textbf{novice:} $\beta_1/\beta_2 > 1+\delta_\rho$\\ 
  $\bullet$ \textbf{expert:} $\alpha/\theta > 1+\delta_r$ \& $\Delta$HbO $\le -\delta_h$};

\node[dec, below=4mm of note] (dec) {In target\\ for $\ge T_p$?\\ (Eq.~\ref{eq:control_law})};

% Side-by-Side Outputs at the Bottom
\node[outbox, below left=6mm and -2mm of dec] (cue) {\textbf{5. Graded cue}\\ haptic/audit./ backing track};

\node[outbox, below right=6mm and -2mm of dec] (hold) {\textbf{Hold}\\ no cue; keep monitoring};

% Connections
\draw[arr] (perf) -- (acq);
\draw[arr] (acq) -- (pre);
\draw[arr] (pre) -- (feat);
\draw[arr] (feat) -- (gate);
\draw[arr] (gate) -- (note);
\draw[arr] (note) -- (dec);

% Decision branches
\draw[arr] (dec) -| node[above left, pos=0.25, font=\tiny] {no} (cue.north);
\draw[arr] (dec) -| node[above right, pos=0.25, font=\tiny] {yes} (hold.north);

% Feedback loops back to Performer (Routing cleanly on outer edges)
\draw[arr] (cue.west) -- ++(-3mm,0) |- node[near start, left, rotate=90, anchor=south, font=\tiny] {closed loop} (perf.west);
\draw[arr, dashed] (hold.east) -- ++(3mm,0) |- (perf.east);

\end{tikzpicture}
\caption{{\textbf{Closed-loop sensing and feedback architecture for musical improvisation.} Signals from a wearable EEG--fNIRS headset are synchronized, cleaned, and reduced to three baseline-normalized features; the expertise gate $\hat E$, set from a behavioral skill set, selects the target set of Eq.~\ref{eq:target}. A cue is delivered only if the state has stayed out of the target for a dwell time $T_p$; otherwise the system holds. Margins $\delta$ are calibrated per performer.}}
\label{fig:bci_pipeline}
\end{figure}

\subsection{{An Expertise-Gated Neurofeedback Trial for Musicians}}
\label{sec:trial}

{The design principle is that the target of intervention should depend on expertise, because stimulation and training effects reverse with expertise \cite{Rosen2016} and because neurofeedback protocols benefit players differently by skill level (A/T for advanced players, SMR for novices \cite{Gruzelier2014, Gruzelier2014b}); experience is also associated with reduced frontoparietal executive activity \cite{pinho2014connecting}. We propose a two-phase trial in musicians (see Table~\ref{tab:expertise_protocol}). We write A/T for the protocol and $\alpha/\theta$ for the marker. \textit{Phase 1 (internalization)} combines deliberate practice with SMR-type training that raises low beta relative to high beta ($\beta_1/\beta_2$) at Cz, aimed at the lower-order attentional and sensorimotor processes that SMR training benefits in novices \cite{Gruzelier2014}. \textit{Phase 2 (relaxation)} adds A/T training (theta over alpha, eyes closed, at Pz \cite{Gruzelier2014b}) and in-performance closed-loop cueing toward $\mathcal{T}_{\text{expert}}$ (as discussed in \secref{sec:control_formulation}).

\textit{Expertise gating.} The gate is a behavioral skill set with blinded expert ratings. The resting $\beta_1/\beta_2$ ratio is deliberately \textit{not} used as a gate, because it is also the Phase-1 training target and would classify any trained novice as an expert. Phase 2 entry requires both the behavioral gate and a stable Phase-1 learning index. The design randomizes participants to expertise-gated training, yoked sham feedback with the same practice, or practice as usual, with blinded quality ratings; the primary neural check is whether Phase-1 attainment is accompanied by lower prefrontal engagement at matched task difficulty, which would support the sequencing and, if absent, would argue against it. Extension to clinical populations (e.g., the autism trial of \secref{sec:beyond} \cite{Bieleninik2017}) is untested and would require its own trial.}

\begin{table*}[ht]
\centering
\caption{{\textbf{Expertise-gated neurofeedback protocol.} Thresholds are baseline-relative design parameters, not literature-derived constants, calibrated per participant. A/T: theta-over-alpha training, eyes closed.}}
\label{tab:expertise_protocol}
{
\footnotesize
\renewcommand{\arraystretch}{1.12}
\begin{tabular}{|p{2.2cm}|p{3.0cm}|p{2.9cm}|p{3.0cm}|p{2.6cm}|p{2.4cm}|}
\hline
\textbf{Phase} & \textbf{Population / gate} & \textbf{Neural target} & \textbf{Protocol} & \textbf{Success metric} & \textbf{Primary outcome} \\
\hline
Phase 1 (internalization) & All novices (below behavioral gate) & $\uparrow\beta_1/\beta_2$ at Cz relative to baseline (sensorimotor gating) & SMR-type training plus deliberate practice & $\beta_1/\beta_2$ above own baseline by calibrated margin, sustained over $\ge$3 sessions; learning index & Behavioral skill rating; DLPFC OxyHb at matched difficulty \\
\hline
Phase 2 (relaxation) & Passed behavioral gate and Phase 1 learning index & $\alpha/\theta$ at Pz above own baseline and $\Delta h\le-\delta_h$ during performance & A/T training \cite{Gruzelier2014b}; closed-loop cueing (Eq.~\ref{eq:control_law}) & Time-in-target fraction; A/T learning index & Blinded expert-rated improvisation quality \\
\hline
\end{tabular}}
\end{table*}

\subsection{Closed-Loop Improv Assistants}

A longer-term goal is a closed-loop engine that monitors a performer's state and supports a target state, a passive BCI using spectral and hemodynamic markers as real-time indicators. While BCIs have focused on motor restoration or command control, a nascent subfield of \textit{passive BCI} for artistic expression has used real-time alpha/theta feedback to modulate generative music \cite{Yuksel2015, Ehrlich2019}. We envision \textit{closed-loop improv assistants} that monitor the performer's state (e.g., detecting hyper-frontal blocking) and deliver subtle cues or simplify environmental constraints (e.g., the backing track) to nudge the performer back toward the target set (Fig.~\ref{fig:bci_pipeline}). {Closed-loop neurofeedback is an established concept \cite{Sitaram2017}, but its in-performance use remains unvalidated (\secref{sec:music_causal}), and continuous neural-state monitoring of performers raises consent and privacy questions that any deployment must address.}

\section{Improvisation Beyond Music}
\label{sec:beyond}

{Outside music, improvisation research is almost entirely behavioral, qualitative, psychophysiological, or simulation-based (Table~\ref{tab:disproportion}). Because neural recordings are rare in these domains, we summarize behavioral, psychophysiological, and early neural findings as reported and note which questions remain open. We begin with the study designs that dominate this literature, so that readers can judge what kind of conclusion each can support.

\subsection{Prevailing Methodology}
\textit{Behavioral experiments} compare improvised with scripted or constrained conditions, or compare experts with novices, using ratings by trained judges, validated questionnaires (e.g., creative self-efficacy, tolerance of uncertainty), or quantitative movement and synchronization measures such as motion-derived complexity and interpersonal coordination in dance and the mirror game \cite{issartel2017understanding, Gueugnon2016}. \textit{Intervention studies} range from randomized trials, e.g., improvisational music therapy \cite{Bieleninik2017}, to pilot and pre--post designs of acting or improv courses \cite{Tang2020, Sutherland2024, Felsman2019}, with outcomes such as symptom scales, social measures, and stress responses. \textit{Qualitative studies} rely on interviews, document analysis, field observation, and case studies, analyzed by thematic coding or systems-based methods \cite{Sasongko2024, Trotter2018, Shafie2024, Hanley2007, miner2001organizational}; \textit{simulation and role-play} studies place trainees in scripted scenarios, as in drama-based training of medical trainees reviewed in \cite{LandryWegener2023}; and \textit{scoping reviews} map the field \cite{kauhanen2020creativity}. These designs offer ecological richness, access to real-world and high-stakes settings, and outcome measures that matter to practitioners. Their typical limits are subjective ratings, small or self-selected samples, heterogeneous constructs, and the absence of any neural measurement, so their results are informative about behavior and outcomes but cannot by themselves identify mechanisms.}

\subsection{Dance and Movement}
Dance improvisation has been studied behaviorally and qualitatively. Expert dancers display greater movement complexity, efficiency, and adaptability in solo and joint improvisation than novices \cite{issartel2017understanding}, and the mirror game shows that even novices improve socio-motor synchronization and movement richness with practice \cite{Gueugnon2016}. Dance improvisation supports real-time sense-making in interdisciplinary workshops \cite{Nelson2024}. The only neural data identified are EEG alpha-activity differences between professional dancers and novices during creative-thinking tasks \cite{Fink2009b} and neurofeedback studies in which alpha/theta plus heart-rate-variability training improved creativity and reduced state anxiety in young dancers \cite{Gruzelier2014d}; neither recorded brain activity during improvised dance.

\subsection{Theater and Social Cognition}
Theatre-based improvisation elicits responses akin to real-life social interaction in a safe, simulated context: improvised fictional and real social rejection produced similar EEG alpha-asymmetry changes \cite{Seppanen2021}, and improvisation training altered behavioral, neuroendocrine, and psychophysiological responses to a social stressor in student teachers \cite{Seppanen2020}. A 12-week acting course for older adults enhanced interaction, social motivation, and inclusion \cite{Sutherland2024}, and improv training improved creative and work-related self-efficacy \cite{Berkemeyer2025}. In psychotherapy, structured improvisation exercises are being explored \cite{Stein2024}.

\subsection{Clinical Applications}
Short-form comedy improvisation changed frontal EEG connectivity and coherence in adolescents with complex developmental trauma in a single-group pilot \cite{DeMichele2021}. The TIPS program was feasible and improved functional outcomes and empowerment in youths at clinical risk for psychosis \cite{Tang2020}. In music therapy, dual-EEG hyperscanning of a therapist--client pair showed inter-brain coupling during musical synchrony \cite{Maidhof2023}, but the large TIME-A randomized trial found that improvisational music therapy in children with autism spectrum disorder fostered engagement without significantly reducing symptom severity \cite{Bieleninik2017}. 
{
Why engagement does not translate into lasting clinical change is unresolved.
}

\subsection{Education}
Brief improvisational exercises increase tolerance for uncertainty and positive affect \cite{Felsman2019, Felsman2020}. Structured improvisation programs in higher education enhance engagement and collaborative problem solving \cite{Rossing2016}, and classroom exercises encouraging unpredictability promote active learning \cite{Berk2009}. In music education, dimensions of improvisation have been characterized by questionnaire \cite{biasutti2017musical}, and process-oriented teaching promotes metacognitive reflection \cite{Biasutti2017}. Situational uncertainty often triggers adaptive responses that rigid planning stifles \cite{sawyer2001creating, nagengast2011uncertainty}.

\subsection{Professional and High-Stakes Settings and Sport}
In healthcare, nurses adapt moment-to-moment to changing patient needs, a skill rarely taught formally \cite{Hanley2007}; drama-based training improved communication and adaptability in medical trainees \cite{LandryWegener2023}. Expertise-driven team improvisation enabled rapid adaptation in the Dutch convalescent plasma project during COVID-19 \cite{Sasongko2024}, and for outdoor activity leaders effective improvisation arose from the interaction of personal expertise, environmental constraints, and organizational factors \cite{Trotter2018}. Improvisation is not uniformly beneficial: well-intentioned unplanned interventions in road safety in Dhaka created hazards \cite{Shafie2024}. These studies illustrate expertise-driven flexibility and rapid adaptation \cite{crossan1996organizational, miner2001organizational}, and its risks without structure. Creativity and improvisation in sport have been reviewed behaviorally \cite{kauhanen2020creativity}.

\subsection{{What Evidence Is Missing}}
{The neural questions that follow from the corpus, and that existing data cannot answer, are: whether the prefrontal findings from musical improvisation appear in dance, theater, or applied improvisation at all; whether novices and experts differ under matched improvisation versus constrained-performance contrasts; how interaction changes neural activity beyond the few hyperscanning case studies; and the time-course of spectral and hemodynamic markers at the onset and offset of improvisation. Answering them requires mobile recording (following the procedure mentioned in \secref{sec:instrumentation}) rather than assumptions transferred from music.}

\section{Conclusion}
\label{sec:conclusion}

{We surveyed 111 studies on the neuroscience of improvisation and found that neurophysiological evidence is concentrated in musical improvisation, while theater, dance, and applied improvisation are covered almost entirely by behavioral and qualitative work. This is a property of the literature, and the survey is organized to reflect it. For music, the studies agree that expertise modulates frontal engagement and that frontal alpha/theta power increases, and they disagree on the direction of lateral prefrontal activity across fMRI, fNIRS, and EEG, on hemispheric lateralization, and on how to interpret alpha power; the dissociated-prefrontal account (transient hypofrontality) is supported by some expert fMRI studies and qualified by others. The current evidence therefore does not point to a universal neural signature of improvisation. Building on the expert-musician evidence, we outlined an expertise-gated closed-loop BCI with explicit provisional targets, instrumentation and acceptance tests, and a neurofeedback trial design. Its value will depend on calibration studies that can confirm or revise the targets, and on mobile recordings in dance, theater, and professional settings that can show which, if any, findings from music transfer.}

\footnotesize{
\bibliographystyle{IEEEtran}
\bibliography{references2}}

\vfill\break
\end{document}